# Albert Einstein and the Fizeau 1851 Water Tube Experiment

Galina Weinstein[*]

In 1895 Hendrik Antoon Lorentz derived the Fresnel dragging coefficient in his theory of immobile ether and electrons. This derivation did not explicitly involve electromagnetic theory at all. According to the 1922 Kyoto lecture notes, before 1905 Einstein tried to discuss Fizeau's experiment "as originally discussed by Lorentz" (in 1895). At this time he was still under the impression that the *ordinary* Newtonian law of addition of velocities was unproblematic. In 1907 Max Laue showed that the Fresnel dragging coefficient would follow from a straightforward application of the *relativistic* addition theorem of velocities. This derivation is mathematically equivalent to Lorentz's derivation of 1895. From 1907 onwards Einstein adopted Laue's derivation. When Robert Shankland asked Einstein how he had learned of the Michelson-Morley experiment, Einstein told him that he had become aware of it through the writings of Lorentz, but only after 1905 had it come to his attention. "Otherwise", he said, "I would have mentioned it in my paper". He continued to say that the experimental results which had influenced him most were stellar aberration and Fizeau's water tube experiment. "They were enough". Indeed the famous Michelson-Morley experiment is not mentioned in the 1905 relativity paper; but curiously Einstein did not mention Fizeau's experimental result either, and this is puzzling in light of the importance of the experiment in Einstein's pathway to his theory. In this paper I try to discuss this question.

## 1. Fizeau's Water Tube Experiment of 1851

In 1818 Augustin Fresnel explained that the motion of the earth does not have any influence on the laws of refraction because *the ether is partially carried along by the earth and light waves inside the optical medium are partially dragged along with the ether*. Fresnel gave a formula for the argument of this partial drag, a formula known by the name "Fresnel's Formula",[1]

$c' = c/n + v(1 - 1/n^2)$,

where c' is the velocity of the light in the medium relative to the earth, v is the velocity of the medium, c is the velocity of light in vacuum, and n is the refractive index.

Fresnel's predictions were confirmed in 1851 by the measurements of Hippolyte Fizeau on the velocity of light in moving water, an experiment which is usually referred to as the water tube experiment.

---

[*] Written while I was at The Center for Einstein Studies, Boston University

On September 29, 1851, Fizeau presented to the Academy of Sciences in Paris, a paper – "Sur les Hypothèthes relatives à l'éther lumineux, et sur une experience qui paraît démontrer que le mouvment des corps change la vitesse avec laquelle la lumière se propage dans leur intérieur" – in which Fizeau confirmed Fresnel's formula, *not the mechanical explanation given by Fresnel to this hypothesis*.[2]

Fizeau had begun his notable paper by saying, "Many theories have been proposed with view of accounting for the phenomenon of aberration of light according to the undulatory theory. In the first instance Fresnel, and more recently Doppler, Stokes, Challis, and several others have published important researches on the subject".[3]

Fizeau endeavored to decide among the different theories. All these hypotheses could be reduced to three, having reference to the state in which the ether "ought to be considered as existing in the interior of a transparent body":

1) "The ether [..,] is fixed to the molecules of the body, and consequently shares all the motions of the body".

Between 1845 and 1846 George Gabriel Stokes proposed a theory in which he assumed that a medium completely carries along the ether within it.[4] Therefore, light will propagate with respect to that medium in the same way as if the medium were at rest; and its velocity of propagation in a transparent refracting medium at rest will also be the velocity of propagation of light with respect to the moving medium. This contradicts Fresnel's *hypothesis*, according to which light is only partially, but not fully, carried along by moving water. Nevertheless Stokes endeavored to obtain Fresnel's *dragging coefficient* and failed to do so from his theory.

Stokes suggested a curious explanation for aberration; he believed that he could reconcile Fresnel's two kinds of ether, the stationary ether and the portion of the ether that is carried along by transparent bodies. According to Stokes, the ether in the earth's vicinity is carried along by all matter (partially by transparent bodies according to Fresnel's theory and fully by non transparent bodies), whereas the ether is at rest at a great distance from earth. This was most peculiar, since Fresnel's formula entailed that light waves were partially carried along by the moving refracting body, and this could by no means correspond to the hypothesis of the ether being completely carried along. Thus Stokes theory was soon plagued by severe problems.[5]

Two other hypotheses mentioned by Fizeau:

2) "The ether is free and independent, and consequently is not carried with the body in its movements". Or:

3) "Only a portion of the ether is free, the rest being fixed to the molecules of the body and, alone, sharing its movements".[6]

The last hypothesis was proposed by Fresnel but it was "far from being considered at present as an established truth, and the relations between ether and matter are still considered, by most, as unknown". Although Fresnel's mechanical explanation given to his hypothesis, "has been regarded by some as too extraordinary to be admitted without direct proofs"; and others simply consider the hypothesis to be at variance with experiment.[7]

Fizeau proposed "an experiment the result of which promised to throw light on the question".[8] Fizeau's "method of observation" was "capable of rendering evident any change of velocity due to motion. It consists in obtaining interference bands by means of two rays of light after their passage through two parallel tubes, through which air or water can be made to flow with great velocity in opposite directions".[9]

The experimental set-up consisted of tubes, built of glass, 5.3 millimeters in diameter. These were traversed by light along their centers, and not near their sides. The lengths of the tubes were quite large, 1.487 meters. Fizeau was worried that difference of temperature or pressure between the two tubes would give rise to considerable displacement of the interference bands, and thus would mask the displacement due to motion. He thus caused the two rays to return towards the tubes by means of a telescope carrying a mirror at its focus. In this mirror each ray was obliged to traverse the two tubes successively; so that the two rays have travelled over exactly the same path in opposite directions. Any effect due to differences of pressure or temperature must necessarily have been eliminated by compensation.

Fizeau placed the two slits quite far apart from each other, but the light was very feeble at the point where interference bands were produced. Hence he placed a convergent lens behind the two slits; the bands were then observed at the point of confluence of the two rays, where the intensity of light was very considerable.

Inside the tubes was water flowing and Fizeau used solar light, admitted laterally and directed towards the tubes by means of reflection from translucent mirror. After their double journey through the tubes, the rays returned and traversed the mirror before reaching the place of interference, where the bands were observed by means of a graduated eye-piece.[10]

In 1903 Michelson wrote that Fresnel gave Fizeau the idea to place the mirror,[11]

"The arrangement of the apparatus which was used in the experiment is shown in Fig. 105. The light starts from a narrow slit S, is rendered parallel by a lens L, and separated into two pencils by apertures in front of the two tubes TT, which carry the column of water. Both tubes are closed by pieces of the same plane-parallel plate of glass. The light passes through these two tubes and is brought to a focus by the lens in condition to produce interference fringes. The apparatus might have been arranged in this way but for the fact that there would be changes in the position of the interference fringes whenever the density or temperature of the medium changed; and, in particular, whenever the current changes direction there would be produced alterations in length and changes in density; and these exceedingly slight differences are quite sufficient to account for any motion of the fringes. In order to avoid this disturbance,

Fresnel had the idea of placing at the focus of the lens the mirror 3f, so that the two rays return, the one which came through the upper tube going back through the lower, and vice versa for the other ray. In this way the two rays pass through identical paths and come together at the same point from which they started. With this arrangement, if there is any shifting of the fringes, it must be due to the reversal of the change in velocity due to the current of water. For one of the two beams, say the upper one, travels with the current in both tubes; the other, starting at the same point, travels against the current in both tubes. Upon reversing the direction of the current of water the circumstances are exactly the reverse: the beam which before traveled with the current now travels against it, etc".

Fizeau concluded from the observed displacement in the interference bands, [12]

"I had already proved that *the motion of air produces no appreciable displacement of the bands*. […]

For water there is an evident displacement. *The bands are displaced towards the right when the water recedes from the observer in the tube at his right, and approaches him in the tube on his left*.

*The displacement of the bands is towards the left when the direction of the current in each tube is opposite to that just defined*.

During the motion of the water the bands remain well defined, and move parallel to themselves, without the least disorder, through a space apparently proportional to the velocity of the water".

After observing displacement of interference bands, Fizeau estimated the observed magnitude of displacement of the interference bands, [13]

"After establishing the existence of the phenomenon of displacement, I endeavored to estimate its magnitude with all possible exactitude". In order to do so, Fizeau varied parameters of the experiments ("magnification of the bands", "the velocity of the water", and so on), so as to be able to measure the magnitude of the displacement. Fizeau calculated the magnitude of the displacement of the interference bands for water velocity of 7.059 meters per second (the observations were made with this velocity), and arrived at the mean value of 0.23 for the displacement of the interference bands.

Next, Fizeau compared the *observed* displacement of the interference bands, 0.23, with that which would result from *theoretical* considerations: first with that which results from the first hypothesis (which is equivalent to Stokes' hypothesis, a mobile ether), and then with that which results from the third hypothesis (partial ether drag – Fresnel's hypothesis). Fizeau did not consider the second hypothesis (an immobile ether), "As to the second hypothesis, it may be at once rejected; for the very existence of the displacements produced by the motion of water is incompatible with the supposition of an ether perfectly free and independent of the motion of bodies".[14]

Fizeau starts with the first hypothesis of the mobile ether. Fizeau defines:

v is the velocity of light in a vacuum, v' the velocity of light in water when at rest,

u the velocity of the water supposed to be moving in a direction parallel to that of the light.

It follows that: v' + u is the velocity of light when the ray and the water move in the same direction,

and v' – u when they move in opposite directions.

Fizeau calls $\Delta$ the required retardation carried along its length,

and E the length of the column of water traversed by each ray.

Fizeau then wrote:

$\Delta = E[v/(v' - u) - v/(v' + u)]$,

or: $\Delta = 2Euv^2/v(v'^2 - u^2)$.

Since u is very small, this expression reduces in the first order to the following expression:

$\Delta = 2Euv^2/vv'^2$.

If m = v/v' is the index of refraction of water, we arrive approximately at:

$\Delta = 2Em^2u/v$.

Since each ray traverses the tubes twice, the length E is double the actual length of the tubes. Calling the latter L = 1.4875 meters, the preceding formula becomes:

$\Delta = 4Lm^2u/v$.

And Fizeau calculated the numerical value and obtained for $\Delta$ the following:

$\Delta = 0.0002418$ millimeters.

Fizeau concluded that this is "the difference of path" which ought to exist between the two rays with reference to the vacuum.[15]

Fizeau then divided the above result by the wavelength. He used the following wavelength $\lambda = 0.000526$, because "the rays about it appear to preserve the greatest intensity after the light has traversed a rather considerable thickness of water". Selecting this ray, then, we find for the displacement the value:

$\Delta/\lambda = 0.4597$.

Recall that previously Fizeau calculated from his *experiment* a displacement of 0.23.[16]

There is no agreement between theory and Fizeau's experiment. Fizeau thus concluded that hypothesis 1) is in conflict with his experiment and went on to the third hypothesis, Fresnel's hypothesis for ordinary phenomena of refraction; in the interior of a body light is propagated with less velocity than in a vacuum. Fresnel supposed that this change in velocity occurs because the density of the ether within a body is greater than that in a vacuum. For two similar media that differ only in their densities, the squares of the velocities of propagation are inversely proportional to these densities: $D'/D = v^2/v'^2$,

D and D' being the densities of the ether in a vacuum and in the body, and v and v' the corresponding velocities. Therefore, $D' = Dv^2/v'^2$, and:

$D' - D = D(v^2 - v'^2)/v'^2$,

The latter gives the excess of density of the interior ether.

According to Fresnel, when the body is put in motion, only part of the interior ether is carried along with it. This part causes the excess in the density of the interior over the surrounding ether. The density of this movable part is: $D' - D$. The other part which remains at rest during the body's motion has the density D.

Fizeau enquired as to the velocity of propagation of the waves in a medium that constituted of an immovable and a movable part, when one supposes the body to be moving in the direction of the propagation of the waves. Fizeau then applied Fresnel's theory to his experimental set-up.[17]

If m is the index of refraction, and L = ½E the length of each tube, one arrives approximately at:

$\Delta = 4L(u/v)(m^2 - 1) = 4L(4/N)n^2(1 - 1/m^2)$.

By inserting the values of the parameters this gives:

$\Delta = 0.00010634$ millimeters.

On dividing this by the wavelength λ of the wave used by Fizeau, the magnitude of the displacement of bands becomes:

$\Delta/\lambda = 0.2022$ millimeters.

The observed value by Fizeau in his experiment was 0.23.

Fizeau concluded, "These values are almost identical".[18] Thus Fizeau found an excellent agreement *between Fresnel's dragging coefficient and his experiment*.

It is important to notice that Fizeau had found an *observable displacement in the interference bands*, and this observable displacement agreed with the theoretical displacement predicted *by Fresnel's formula*. As to partial ether drag, *Fizeau did not observe any moving ether, and he did not approve the "partial ether" drag hypothesis*.

## 2. Lorentz Derives Fresnel's Dragging Coefficient in his 1892 Electron Theory

In his 1892 paper, *La théorie électromagnétique de Maxwell et son application aux corps Mouvants*, Lorentz advanced a version of the theory of the electron, based on Maxwell's electromagnetic theory in order to explain electromagnetic, as well as optical, phenomena in bodies at rest and in motion. Lorentz demarcated ponderable matter from the imponderable luminiferous *stationary immobile* ether. In section §160 of this paper, "Entrainement des Ondes lumineuses par la matière pondérable" (Entrainment of light waves in ponderable matter) Lorentz derived Fresnel's dragging coefficient from his interpretation of Maxwell's equations.[19]

*Lorentz interpreted Fresnel's formula in such a way that it is the waves that are partially dragged by the dielectric medium and not the ether.* This was in fact Fizeau's original interpretation of Fresnel's result in his 1851 paper as discussed above.

According to Lorentz, when a homogenous dielectric is at rest, let the velocity of propagation of the waves be $W_0$. The velocity of propagation of the waves for moving media is,[20]

$$(I) \quad W = \pm W_0 - \frac{p}{v^2},$$

With $v$ is the index of refraction, where $p'$ the component of the velocity of the body in the direction of wave's propagation. This is the velocity of propagation of the waves with respect to the ponderable matter.

The velocity of the waves with respect to the ether is then,[21]

$$(II) \quad \pm W_0 + \left(1 - \frac{1}{v^2}\right)p.$$

Lorentz said that regardless of the direction to which the waves propagate, the fraction that comes before p is always,[22]

$$(158) \quad 1 - \frac{1}{v^2}$$

Lorentz ended his derivation by saying, "Remarquons encore que d'apres notre théorie, la valeur (158) est applicable à chaqué espèce de lumière homogène, si seulement on entend par v l'indice de refraction qui lui est proper".[23]

## 3. Lorentz Derives Fresnel's Dragging Coefficient in his 1895 Electron Theory

Same as 1892 Lorentz derived Fresnel's formula in sections §68 and §69 of the chapter "*Die Mitführung der Lichtwellen durch die ponderable Materie*" (the entrainment of light waves by ponderable matter) of his 1895 *Versuch einer theorie der electrischen und optischen erscheinungen bewegten kõrpern* [equations (79) to (84)]:[24] This derivation does not explicitly involve electromagnetic theory at all. The crux of the matter and of the whole derivation was that Lorentz applied the theorem of corresponding states from his 1895 section §59 to equation (79) below, and obtained the transformed expression for the waves.[25]

Consider a body *at rest*. It is either isotropic or anisotropic. Beams of plane light waves propagating within the body will take the form,

$$(79) \quad A\cos\frac{2\pi}{T}\left(t - \frac{b_x x + b_y y + b_z z}{W} + B\right).$$

W is the velocity of propagation, and $b_x$, $b_y$, $b_z$, are the direction cosines of the wave normal and T is the period = 1/frequency. Now we impart to the body a velocity **p**. "Nachdem man dem Körper eine Geschwindigkeit **p** ertheilt hat, kann, wie wir sahen (§ 59), in demselben ein Bewegungszustand bestehen, für welchen Ausdrücke wie",[26]

$$A\cos\frac{2\pi}{T}\left(t' - \frac{b_x x + b_y y + b_z z}{W} + B\right) = A\cos\frac{2\pi}{T}\left(t - \frac{b'_x x + b'_y y + b'_z z}{W'} + B\right).$$

W' is the velocity with which the waves of *relative* oscillation of period *T* are propagating in the direction $b'_x$, $b'_y$, $b'_z$ within the body *after applying the theorem of corresponding states – local time*.

If $p_n$ is the component of the velocity of the body in the direction of the wave normal, which is related to W', then,

$$(82) \quad W' = W - p_n \frac{W^2}{c^2}.$$

Suppose the body is isotropic. The velocity W is independent of the direction of the waves.

*And thus the index of refraction, c/W = N of the body at rest depends only on T.*

Equation (82) reduces to,

$$(83) \quad W' = W - \frac{p_n}{N^2}.$$

The description of the phenomena until now was for a coordinate system moving together with the ponderable matter. Therefore, (83) represents the velocity of the moving light waves relative to the matter. [This equation is equivalent to equation (I) of Lorentz's 1892 paper, for one direction of propagation]

The relative velocity with respect to the (stationary) ether is the velocity of the light waves (83) with respect to matter in the direction of the wave normal together with the component $p_n$ of translational velocity of the body in this direction; this gives,

$$(84) \quad W'' = W + \left(1 - \frac{1}{N^2}\right) p_n.$$

Lorentz concluded that it agrees with the known Fresnel hypothesis ("was mit der bekannten Annahme Fresnel's übereinstimmt").[27]

After obtaining equation (84), Lorentz went on to derive in sections §70-§71 the relative period T of oscillation of light.[28] Recall that in equation (82) the index of refraction, c/W = N of the body at rest depends only on T. Lorentz derived an extra equation (90) which *deviated from (84) in an extra term*. He used for this derivation the Doppler Effect, the Doppler variation of the period of oscillation T of the light source.

Therefore, Lorentz's innovations in his latter derivation of Fresnel's dragging coefficient were twofold; he first implemented the theorem of corresponding states and the local time to obtain equation (84). After getting equation (84), he corrected this equation by taking into account the Doppler effect. By doing this he actually implicitly called for reinterpretation of the result of the latest Fizeau experiments, which did not take into account the Doppler effect. But with hindsight and from the relativistic perspective, Lorentz connected between Fresnel dragging coefficient, the relativistic addition theorem of velocities, and the Doppler effect.

Could Einstein in his first relativity paper in section §7 "Theory of Doppler's Principle and of Aberration" of his 1905 relativity paper,[29] have been influenced from Lorentz's additional derivation, an extension of equation (84) to Lorentz's equation (90) below using the Doppler Effect? It appears he was not influenced by Lorentz's derivation, probably for the same reason he was not inspired by the previous derivation – implementing the theorem of corresponding states and the local time to obtain equation (84). In his first relativity paper of 1905, Einstein did not apply the relativistic addition theorem of velocities to solve optical problems.

Let us now see how Lorentz corrected equation (84). In Fizeau's experiment, two adjacent glass tubes were used, which were closed by glass plates. Through them the water was flowing with the same velocity **p**, but in opposite directions. The two light beams, which interfered with each other, passed through the apparatus, so that one was propagated in both tubes in the direction of the water stream, and the other in the opposite direction.

Lorentz considered a fixed point $P$ in the interior of one of the tubes. When the water flow is stationary, the conditions, under which the light is propagating from the source to this point, remain the same, and this applies to *both* directions from which the rays reach point $P$. Impulses, coming with certain periods from the source, will arrive at with the same periods, and if $T$ is the period of oscillation of the light source, then it is also the period at $P$. It follows that for the *relative* oscillation period related to the water,

$$\left(1 \pm \frac{p}{W}\right)T.$$

The + or − sign is applied depending on whether the light bundle is propagating in the direction of the water motion or in the opposite direction. W is given by equation (82) and for the body at rest.

The corresponding indices of refraction once the relative period $\pm T'$ is taken into account are,

$$n \pm \frac{p}{W} T \frac{dn}{dT}$$

and,

$$W = \frac{c}{N} \mp \frac{p}{n} T \frac{dn}{dT}$$

By (82) and [$p_n \mp p$], Lorentz found the expression for $\pm W'$,

$$\pm W' = \frac{c}{n} \mp \frac{p}{n^2} \mp \frac{p}{n} T \frac{dn}{dT},$$

And the relative velocity with respect to the ether is, [30]

$$(90) \quad W'' = \frac{c}{n} \mp p\left(1 - \frac{1}{n^2}\right) \mp \frac{p}{n} T \frac{dn}{dT} = \frac{c}{n} \mp p\left(1 - \frac{1}{n^2}\right) + \frac{1}{n} T \frac{dn}{dT}.$$

This formula is somewhat different from (84).

According to Lorentz, [31]

$$\varepsilon = 1 - \frac{1}{n^2} - \frac{1}{n} T \frac{dn}{dT},$$

and Lorentz calculated the value ε = 0, 451. According to Michelson and Morley's 1886 repetition of Fizeau's experiment, ε = 0,434 (or ε = 0.438 including possible statistical errors) for the equation:

$$(91) \quad W = \frac{c}{n} \pm v\left(1 - \frac{1}{n^2}\right),$$

which is equivalent to equation (84).

Lorentz ended by saying, [32]

"Sollte es gelingen, was zwar schwierig, aber nicht unmöglich scheint, experimentell zwischen den Gleichungen (90) und (91) zu entscheiden, und sollte sich dabei die erstere bewähren, so hätte man gleichsam die DOPPLER'sche Veränderung der Schwingungsdauer für eine künstlich erzeugte Geschwindigkeit beobachtet. Es ist ja nur unter Berücksichtigung dieser Veränderung, dass wir die Gleichung (90) abgeleitet haben".

Lorentz proposed to experimentally decide between his equation (90) and equation (91), which is equivalent to (84) [that of Fresnel's]. Lorentz explained that equation (90) was derived by taking into account the Doppler variation of the period of oscillation of an artificially generated velocity; and if his derivation towards (90) was justified then his above suggestion was justified as well.

**4. 1905: Einstein does not mention Fizeau – derives Aberration and Doppler**

On February 4 1950 Robert Shankland met Einstein at his office in Princeton, "When I asked him how he had learned of the Michelson-Morley experiment, he told me that he had become aware of it through the writings of H. A. Lorentz, but *only after 1905* had it come to his attention! 'Otherwise', he said, 'I would have mentioned it in my paper'. He continued to say the experimental results which had influenced him most were the observations on stellar aberration and Fizeau's measurements on the speed of light in moving water. 'They were enough', he said".[33]

Indeed the famous Michelson-Morley second order in v/c ether drift experiment is not mentioned in the 1905 relativity paper; but curiously Einstein did not mention the result of Fizeau's experimental result in his 1905 relativity paper either, and this is puzzling in light of the importance of the experiment in Einstein's pathway to his theory. Einstein only mentions and derives stellar aberration in the relativity paper. Let us try to explain this now.

In section §7 of the relativity paper which discusses the optics of moving bodies Einstein considered the system *K*. Very far from the origin of *K*, there is a source of electromagnetic waves. Let part of space containing the origin of coordinates be

represented to a sufficient degree of approximation by the following equations (plane waves):

$X = X_0 \sin \Phi$, $Y = Y_0 \sin \Phi$, $Z = Z_0 \sin \Phi$,

$L = L_0 \sin \Phi$, $M = M_0 \sin \Phi$, $N = N_0 \sin \Phi$.

The phase is: $\Phi = \omega[t - (ax + by + cz)/c]$
and $a, b, c$ are the direction cosines and wave-normal **n**; $\omega$ is the frequency of the plane waves (really a sphere, but the radius is so big that it can be treated as plane waves).

Einstein asks: What characterizes the waves when they are examined by an observer at the same point 0, but at rest in the system $k$?[34]

Einstein applies the transformations he found in section §3 (the Lorentz transformations) for the coordinates and time, and those for electric and magnetic forces that he found in section §6:

$X' = X$, $Y' = \beta(Y - vN/c)$, $Z' = \beta(Z + vM/c)$,
$L' = L$, $M' = \beta(M + vZ/c)$, $N' = \beta(N - vY/c)$

[(X, Y, Z), (L, M, N) – the electric and magnetic force vectors satisfying a set of Maxwell-Hertz equations, (X', Y', Z') and (L', M', N') – in the moving system $k$ satisfying a set of Maxwell-Hertz equations of the same form as in K; $\beta$ is for $\frac{1}{\sqrt{1-v^2/c^2}}$.], to the equations of the plane electromagnetic wave with respect to $K$. He obtains a new set of equations, from which he deduces new transformation equations for the frequency $\omega$ and direction cosines of the wave normal **n'**, Einstein obtains the Doppler principle and the equation that expresses the law of aberration.

The Doppler principle for any velocity is: "If an observer is moving with velocity v relative to an infinitely distant source of light of frequency ν, in such a way that the connecting line 'light source-observer' forms an angle $\varphi$ with the velocity of the observer, which is referred to the coordinate system at rest relative to the source of light, the frequency ν' of the light perceived by the observer, is given by the equation":[35]

$\nu' = \nu(1 - \cos \varphi \, v/c)/\sqrt{(1 - v^2/c^2)}$.

When $\varphi = 0$ the equation assumes the simple following form:

$\nu' = \nu\sqrt{[(1 - v/c)/(1 + v/c)]}$.

If the angle between the wave-normal (direction of the ray) in the moving system $k$ and the direction of motion is $\varphi'$, the equation for $a'$ gives the law of aberration in its most general form:

$cos\ \varphi' = (cos\ \varphi - v/c)/(1 - cos\ \varphi\ v/c)$.

If $\varphi = \pi/2$ the equation reduces to: $cos\ \varphi' = -v/c$. [36]

Einstein presented in the kinematical part a new addition law for relative velocities, but he did not derive Fizeau's experimental result from this law:

$u_x = (u'_x + v)/[1 + (vu'_x/c^2)]$.

Einstein did not recognize that the result of the Fizeau experiment could be obtained using this law. He did not derive stellar aberration in a similar manner either. In section §7 of the 1905 relativity paper Einstein derived the latter result from the same transformation equations of the wave normal that give the Doppler effect without mentioning the addition law of velocities.

In his 1907 paper, "The Entrainment of Light by Moving Bodies According to the Principle of Relativity", Max Laue obtained Fresnel's coefficient (Fizeau's result) from Einstein's addition theorem of velocities. The result was rather complicated in 1907, and afterwards Laue simplified it when he considered that the direction of the light ray coincides with that of the motion of the observer relative to the medium. Laue considered two coordinate systems with parallel axes. The "primed" system and the "unprimed" system are moving relative to each other along the direction of $X$ with velocity v. A velocity $w$ in respect to the primed system, which direction forms the angle $\theta$ with the $X'$-axis, corresponds to a velocity in the unprimed system:

$w = \sqrt{[v^2 + w'^2 + 2vw'cos\theta' - (1/c^2)v^2w'^2sin^2\theta']}/[1 + (1/c^2)vw'cos\theta']$. [37]

If a body of refractive index $n$ is at rest in the primed system, then the phase velocity of light in the primed system is:

$w' = c/n$.

The corresponding velocity in the unprimed system is therefore,

$w = \sqrt{[v^2 + (c^2/n^2) + 2v(c/n)cos\theta' - (v^2/n^2)\ sin^2\theta']}/[1 + (v/cn)\ cos\theta']$.

If the directions of the velocities v and $c/n$ coincide, as in the experiment of Fizeau, then it is $cos\theta' = \pm 1$ and:

$w = (c/n \pm v)/(1 \pm v/cn) = c/n + (1 - 1/n^2)$ to first order. [38]

In his 1913 paper "Das Relativitätsprinzip", Max Laue put $u'_x = u' = (c/n)$ in:

$u_x = (u'_x + v)/[1 + (vu'_x/c^2)]$.

For an observer in *K'* moving with the medium of refractive index *n*, light is propagated with velocity *c/n* in all directions. Its velocity of propagation as seen by an observer *K* moving with velocity *v* relative to the medium is *q* (and not *c/n* + v as it would be according to an emission theory). Thus, $u_x = u = q$. If the direction of the light ray coincides with that of the motion of the observer relative to the medium, the addition theorem for velocities above of Einstein from 1905 gives:

$q = [v + (c/n)]/[1 + (v/cn)]$.

Since v is small as compared to *c*, when first-order terms only are retained this expression approximately gives:

$c/n + v(1 – 1/n^2)$,

Which is Fresnel's formula.[39]

In 1907, Laue wrote Einstein, "Upon my return I came across an article in the *Annalen* by J. Laub, 'Optics of Moving Media',[40] which derives the dragging coefficients from the relativity principle. The paper has a very neat basic idea, but, unfortunately, it later on contains two direct errors and one inconsistency in the reasoning, namely an interchanging of group and phase velocity. Even so, I did not withdraw my derivation; rather, it will appear in one of the next issues.[41] I have communicated my objections to the author".[42]

Laue's derivation is mathematically equivalent to Lorentz's derivation of 1895. From the relativistic perspective Lorentz's derivation shows that a wave propagating through some medium in the x-direction with velocity W for a co-moving observer (using time coordinate t') has, in first order approximation, velocity (84) for a second observer (using time coordinate t) with respect to whom the first is moving with velocity $p_n$ in the x direction, where t and t' are interpreted as the ordinary times in each system.[43]

Before 1905 Einstein indeed tried to discuss Fizeau's experiment "as originally discussed by Lorentz" in 1895. According to the 1922 Kyoto lecture notes Einstein said:[44] "Then I tried to discuss the Fizeau experiment on the assumption that the Lorentz equations for electrons should hold in the frame of reference of the moving body as well as in the frame of reference of the vacuum as originally discussed by Lorentz". However, Einstein was still under the impression that the ordinary Newtonian law of addition of velocities was unproblematic.

John Norton writes, "It might seem surprising that Einstein could devise and publish the relativistic rule of velocity composition in his 1905 paper (§5) without recognizing that the result of the Fizeau experiment is a vivid implementation of the rule." And: "The situation with stellar aberration is similar. The result can be arrived at rapidly by means of the relativistic rule of velocity composition. Yet Einstein (1905, §7) derives the result from the same transformation of the waveform that gives the Doppler shift without mention of velocity composition".[45]

Stachel explains that Einstein was under the spell that everything could be solved using Maxwell's equations that he failed to notice the kinematic nature of Fresnel's formula, resulting from direct application of the relativistic law of combination of relative velocities; it was left for Laue to make this observation in 1907.[46]

In 1905 Einstein wanted to show in a heuristic way that aberration and cognate optical phenomena could be derived from the transformation of the waveform and *not* derived from the relativistic addition law of velocities. He wanted to demonstrate that the optics of moving bodies problems could be solved using his new kinematics (§3) – the principle of relativity and the light postulate and the Maxwell equations. Einstein said in his relativity paper that all problems in the optics of moving bodies could be solved by the heuristic method employed here.[47]

In order to derive Fizeau's result from the transformation of the wave he probably needed a complicated calculation. It appears that he could not derive in a simple manner the Fizeau result from the transformation of the waveform. In later papers he followed a different path and derived aberration and Fizeau's result by means of the relativistic addition law of velocities – following Laue's derivation.

Tel-Aviv, 2012.

*I wish to thank Prof. John Stachel from the Center for Einstein Studies in Boston University for sitting with me for many hours discussing special relativity and its history. Almost every day, John came with notes on my draft manuscript, directed me to books in his Einstein collection, and gave me copies of his papers on Einstein, which I read with great interest.*

---

[1] Fresnel, Augustin 1818 "Lettre d'Augustin Fresnel à Franscois Arago sur l'influence du mouvement terrestre dans quelques phénomènes d'optique", *Annales de chimie et de physique* IX, 57 (1818); Reprinted in Fresnel, *Oeuvres Complètes 1866-1870*, Paris: Imprimerie impérial, Vol II, pp. 627-636; pp. 634-636.

[2] Fizeau, Armand-Hippolyte, "Sur les hypotheses relatives a l'ether lumineux, et sur une experience qui parait demontrer que le mouvement des corps change la vitesse avec laquelle la lumiere se propage dans leur interieur", *Comptes Rendus de l'Académie des Sciences* 33, 1851, pp. 349-355.

[3] Fizeau, 1851, p. 349; Fizeau, Armand-Hippolyte, "On the Effect of the Motion of a Body upon the Velocity with which it is Traversed by Light", *Philosophical Magazine and Journal of Science* 19, April 1860, pp. 245-260. (English translation of the 1859 paper published in *Annales de Physique et chimie* 57, pp. 385-404), p. 245 (The quotations are taken from the official English translation).

[4] Stokes, George Gabriel, "On the Aberration of Light", *Philosophical Magazine* 27, 1845, pp. 9-15; Stokes, George Gabriel, "On Fresnel's Theory of the Aberration of Light", *Philosophical Magazine* 28, 1946, pp. 76-81.

[5] Stokes, 1846, p.147.

[6] Fizeau, 1851, p. 349; Fizeau, 1860, p. 245.

[7] Fizeau, 1851, p. 350; Fizeau, 1860, p. 246.

[8] Fizeau, 1851, p. 350; Fizeau, 1860, p. 246.

[9] Fizeau, 1851, p. 350; Fizeau, 1860, p. 247.

[10] Fizeau, 1851, p. 352; Fizeau, 1860, p. 247.

[11] Michelson, Albert Abraham, *Light Waves and Their Uses*, 1903, Chicago: Chicago University Press, pp. 153-154.

[12] Fizeau, 1851, p. 353; Fizeau, 1860, p. 249.

[13] Fizeau, 1851, p. 353; Fizeau,1860, pp. 249-250.

[14] Fizeau, 1860, p. 251.

[15] Fizeau, 1860, pp. 251-252.

[16] Fizeau, 1860, p. 252.

[17] Fizeau, 1860, pp. 252-253.

[18] Fizeau, 1860, pp. 253-254.

[19] Lorentz, Hendryk Antoon, "La théorie électromagnétique de Maxwell et son application aux corps Mouvants", *Archives néerlandaises des sciences exactes et naturelles* 25, 1892, pp. 363-552; *Extrait des Archives néerlandaises des sciences exactes et naturelles* 25, Leiden: E.J. Brill, 1892, pp. 1-190; reprinted in *Collected Papers 1935-1939*, The Hague: Nijhoff, 9 Vols, Vol. 2, pp. 162-343, pp. 162-164.

[20] Lorentz, 1892, p. 163.

[21] Lorentz, 1892, p. 164.

[22] Lorentz, 1892, p. 164.

[23] Lorentz, 1892, p. 164.

[24] Lorentz, Hendryk Antoon, *Versuch einer Theorie der elektrischen und optischenen Erscheinungen in bewegten Körpern*, 1895, Leiden: Brill, pp. 96-97.

[25] Lorentz, 1895, p. 87.

[26] Lorentz, 1895, p. 96.

[27] Lorentz, 1895, p. 97.

[28] Lorentz, 1895, pp. 100-101.

[29] Einstein, Albert, "Zur Elektrodynamik bewegter Körper, *Annalen der Physik* 17, 1, 1905, pp. 891-921; pp. 910-911.

[30] Lorentz, 1895, p. 101.

[31] Lorentz, 1895, p. 102.

[32] Lorentz, 1895, p. 102.

[33] Shankland, Robert, "Conversations with Albert Einstein I/II" *American Journal of Physics* 31, 1963, pp. 47-57; 41; p. 48.

[34] Einstein, 1905, pp. 910-911.

[35] Einstein, 1905, p. 911.

[36] Einstein, 1905, pp. 911-912.

[37] Laue, Max, "Die Mitführung des Lichtes durch bewegte Körper nach dem Relativitätsprinzip", *Annalen der Physik* 23, 1907, pp. 989–990; p. 989.

[38] Laue, 1907, p. 990.

[39] Laue, von Max, "Das Relativitätsprinzip", *Jahrbücher der Philosophie* 1, 1913, pp. 99-128 in Laue, Laue von Max, *Gesammelte Schriften und Vorträge*, 1961, Berlin: Friedr. Vieweg & Sohn, p. 264; Pauli, Wolfgang, *Theory of Relativity*, 1958, Oxford and New York: Pergamon, p. 17-18; Pauli, Wolfgang, "Relativitätstheorie" in *Encyklopädie der Mathematischen wissenschaften*,Vol. 5, Part 2, 1921, Leipzig: Teubner, 2000, pp. 32-34 (pp. 562-564).

[40] Laub, Jacob, "Zur Optik der bewegten Körper", *Annelen der Physik* 328, 1907, pp. 738–744.

[41] Laue, 1907.

[42] Laue to Einstein, September 4, 1907, *The Collected Papers of Albert Einstein. Vol. 5: The Swiss Years: Correspondence, 1902–1914*, Klein, Martin J., Kox, A. J., and Schulmann, Robert (eds.), Princeton: Princeton University Press, 1993, Doc 57.

[43] Janssen Michel and Stachel, John, "The Optics and Electrodynamics of Moving Bodies", *Max Planck Institute for the History of Science*, preprint 265, 2004, p. 34.

[44] Einstein, Albert, "How I Created the Theory of Relativity, translation to English by. Yoshimasha A. Ono, *Physics Today* 35, (talk – 1922) 1982, pp. 45-47; p. 46.

[45] Norton, John, "Einstein's investigations of Galilean covariant Electrodynamics prior to 1905", *Archive for the History of Exact Sciences* 59, 2004, pp. 45-105; p. 93.

[46] Stachel, John, "Fresnel's (Dragging) Coefficient as a Challenge to 19th Century Optics of Moving Bodies", in Eisenstaedt Jean and Kox A. J., *The Universe of General Relativity. Einstein Studies*, vol. 11, 2005, Boston: Birkhäser, pp. 1-13; p. 10.

[47] Einstein, 1905, p. 916.